\documentclass[twocolumn,aps,prl]{revtex4}

\usepackage{amsmath,epsfig,amsfonts,epsfig,graphicx,rotating}

\begin{document}

\title{\bf Trapping and transmission of matter-wave solitons in a collisionally inhomogeneous environment}

\author{
G.\ Theocharis$^{1}$,
P.\ Schmelcher$^{2,3}$, 
P. G.\ Kevrekidis$^{4}$ and 
D. J.\ Frantzeskakis$^{1}$ 
}
\affiliation{
$^{1}$ Department of Physics, University of Athens, Panepistimiopolis,Zografos, Athens 157 84, Greece \\ 
$^{2}$ Theoretische Chemie, Physikalisch-Chemisches Institut, Im Neuenheimer Feld 229, 
Universit\"at Heidelberg, 69120 Heidelberg, Germany \\
$^{3}$ Physikalisches Institut, Philosophenweg 12, Universit\"at Heidelberg, 69120 Heidelberg, Germany \\
$^{4}$ Department of Mathematics and Statistics,University of Massachusetts, Amherst MA 01003-4515, USA 
}
\begin{abstract}
We investigate bright matter-wave solitons in the presence of a spatially varying scattering length.
It is demonstrated that, even in the absence of any external trapping potential,
a soliton can be confined due to the inhomogeneous collisional interactions.
Moreover we observe the enhanced transmission of matter-wave solitons through 
potential barriers for suitably chosen spatial variations of the scattering length.
The results indicate that the manipulation of atomic interactions can become a versatile tool
to control the dynamics of matter waves.
\end{abstract}

\maketitle 



{\it Introduction.} The recent developments in the field of 
atomic Bose-Einstein condensates (BECs) \cite{dalfovo} has inspired, among others, 
many studies on their nonlinear excitations. Especially, as far as matter-wave solitons 
are concerned, dark \cite{dark}, bright \cite{expb1,expb2} and gap \cite{gap} solitons have 
been observed experimentally and studied theoretically. 
Atom optical devices such as the atom chip \cite{Folman} offer the possibility to control
and manipulate matter-wave solitons. Their formal similarities with optical solitons
indicate that they may be used in applications similarly to their optical counterparts \cite{kivpr}. 

Typically bright (dark) matter-wave solitons are formed in BECs with attractive (repulsive)  
interatomic interactions, i.e., for atomic species with negative (positive) scattering length $a$.  
Employing magnetically-induced Feshbach resonances both the magnitude and sign of the
scattering length can be changed by tuning the external magnetic field (see 
e.g.  \cite{feshbachNa}
and also \cite{expb1,expb2} where the Feshbach resonance in $^{7}$Li condensates was used 
for the formation of bright matter-wave solitons). 
These studies paved the way for important experimental discoveries, such as the formation 
of molecular BECs \cite{Herbig03} and the revelation of the BEC-BCS crossover \cite{Bartenstein04}.
From the theoretical viewpoint, it was predicted that a time-dependent modulation of the 
scattering length can be used to prevent collapse in higher-dimensional attractive BECs \cite{FRM1}, 
or to create robust matter-wave breathers \cite{FRM2}. Adding to a constant bias magnetic field
a gradient in the vicinity 
of a Feshbach resonance allows for a spatial variation of the
scattering length over the ensemble of cold atoms thereby yielding a collisionally inhomogeneous
condensate. Due to the availability of magnetic and optical (laser-) fields the external trapping
potential and the spatial variation of the scattering length can be adjusted independently.
Moreover hyperfine species with the magnetic quantum number $M_F=0$ do not feel a potential due to the magnetic field
but experience magnetically-induced Feshbach resonances (see e.g. \cite{Blume}).
In this case the external potential is formed exclusively by an optical dipole potential
and the magnetic field configuration is responsible for the spatially dependent scattering
length. Recently this has been exploited to study the properties of cold atomic gases in a collisionally
inhomogeneous environment (CIE) \cite{fka,our}.

Here we investigate the dynamics of bright matter-wave solitons of a quasi one-dimensional
(1D) $^{7}$Li BEC \cite{expb1,expb2} in a CIE demonstrating the appearance of unexpected phenomena
which make the spatial manipulation of the scattering length a versatile tool.
Firstly we show that the inhomogeneity of $a$ induces an effective confining potential felt by a
bright matter-wave soliton, even in the absence of any external trapping potential.
The proposed scheme leads to a {\it collision-induced breathing soliton}, which oscillates due to the
effective confinement and is periodically compressed when passing through the region of a large scattering length. 
As a second prototypical situation for a CIE, we consider the transmission of a soliton through a potential 
barrier underneath which a suitably chosen spatially varying scattering length $a(x)$ is present.
This setup allows to enhance the transmission of the soliton, i.e., the barrier becomes more transparent
compared to the case of a spatially independent atom-atom interaction. 
More specifically,
we use the particular form $a(B)$ for a $^{7}$Li condensate near a corresponding
Feshbach resonance (see below). This species has already been used in order to prepare bright matter-wave solitons.
We believe that the {\it collision-induced trapping and transmission} of matter-wave solitons reported here,
are two generic phenomena illustrating that collisionally inhomogeneous matter-waves 
exhibit a number of interesting and fundamentally new features that 
could also be relevant for future applications.

{\it Collision-induced trapping.} To specify our setup we choose the magnetic field dependence 
of the scattering length $a$ of a $^{7}$Li condensate provided in Ref.\cite{expb2} (see also Ref. \cite{expb1}).
However, we emphasize that this case serves only as a typical example of such
a curve, used for concreteness.
Let us focus on the regime $0 \le B \le 590$ G, which is far left from the Feshbach resonance at $720$ G, 
where the scattering length is small and the inelastic collisional loss of atoms is practically negligible.
We have $a(B)<0$ for $150$ G $<B<520$ G and $a(B)>0$ elsewhere. At $B \approx 352$ G, the scattering length reaches a minimum
with $a \approx -0.23$ nm. The experimentally observed quasi-1D bright matter-wave solitons
\cite{expb1,expb2} have been created in the above-given regime of negative scattering length.

Let us assume a magnetic field configuration $B=B_{0}+\epsilon x$ (G), 
where $\epsilon$ is the field gradient and $B_{0}=450$ G, being far from the position of the resonance.
The gradient chosen is of the order of a few tens of G/$\mu$m which can be experimentally realized
in microscopic matter-wave devices such as the atom-chip \cite{Folman}. Other resonances and
species might require much smaller values of the gradient to implement a significant change of the
scattering length on the scale of the condensate. To extract the spatial dependence of the 
scattering length we apply a fifth-order polynomial fit to the data for $a(B)$ 
given in Ref. \cite{expb2}.
The result is shown in Fig. 1 for two nonzero values of $\epsilon$. The function $a(x)$ 
possesses a single minimum and the dependence on $x$ 
is, as expected, much more dramatic for
a larger value of $\epsilon$. For $\epsilon =0$ we have $a=-0.182$ nm.

\begin{figure}[tbp]
\includegraphics[width=6cm]{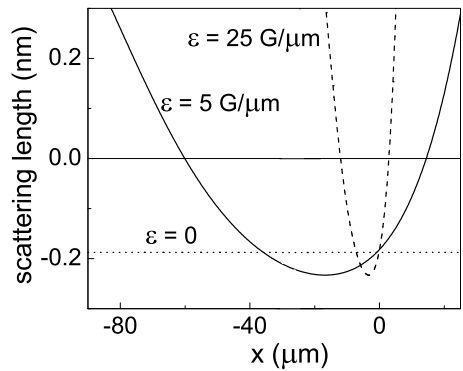}
\caption{The spatial variation of the scattering length for $B=450+\epsilon x$ (G) and magnetic field gradients  
$\epsilon=0$ (dotted line), $\epsilon=5$ G/$\mu$m (solid line) and $\epsilon=25$ G/$\mu$m (dashed line).}
\label{fig1}
\end{figure}

The evolution of an {\it untrapped}, quasi-1D bright matter-wave soliton in a CIE is described by 
the following normalized Gross-Pitaevskii (GP) equation:
\begin{equation}
i \partial_{t} \psi = - \frac{1}{2} \partial_{x}^{2} \psi -g(x)|\psi|^2 \psi. 
\label{GP}
\end{equation}
Here, $\psi$ is the mean field wavefunction (with the density $|\psi|^2$ measured in units of the peak
density $n_0$), $x$ is given in units of the healing length $\xi=\hbar/\sqrt{n_{0} g_{0} m}$
(where $g_{0}=2 \hbar \omega_{\perp} a(B_{0})$ and $\omega_{\perp}$ is the confining frequency
in the transverse direction), and the time unit is $\xi/c$ (where $c=\sqrt{n_{0}g_{0}/m}$ is the
Bogoliubov speed of sound). Finally, the spatially dependent nonlinearity is given by $g(x) = a(x)/a(B_{0})$,
with $a(x) \equiv a\left(B_{0}+\epsilon \xi x\right)$ 
(note $g(x=0)=g(\epsilon=0)=1$).
Typically, for a quasi-1D $^{7}$Li condensate with $\omega_{\perp}= 2\pi \times 1000$Hz and
$n_{0}=10^9$ m$^{-1}$, the healing length and speed of sound amount to $\xi= 2\mu$m and $c=4.6$mm/s, respectively. 

Introducing the transformation $\psi=u / \sqrt{g}$ in the region $g(x)>0$  we reduce Eq. (\ref{GP})
to the following perturbed nonlinear Schr\"{o}dinger (NLS) equation,   
\begin{eqnarray} 
i \partial_{t} u + \frac{1}{2} \partial_{x}^{2} u + |u|^{2} u = R(u),
\label{gpe1d_u} 
\end{eqnarray} 
with a perturbation $R(u) \equiv \frac{d}{dx}\ln(\sqrt{g}) \partial_{x} u + O\left(L_{S}^{2}/L_{B}^{2}\right)$. 
The first term of $R$ is of the order $O\left(L_{S}/L_{B}\right)$, where  
$L_{S}$ and $L_{B}$ are the characteristic spatial scales of the soliton and of the inhomogeneity due to the 
magnetic field gradient, respectively. In the case $R=0$ 
($\Leftrightarrow \epsilon=0$), Eq. (\ref{gpe1d_u}) has a bright soliton solution of the form \cite{zsb},
\begin{eqnarray} 
u(x,t)=\eta {\rm sech}[\eta(x-x_{0})]\exp[i(kx-\phi(t)], 
\label{sol} 
\end{eqnarray} 
where $\eta$ is the amplitude and inverse 
width of the soliton, and $x_{0}$ is the soliton center. 
The parameter $k=dx_{0}/dt$ defines both the soliton velocity and wavenumber, and $\phi(t)=(1/2)(k^2-\eta^2)t$ 
is the soliton phase. For the above mentioned typical values of the parameters, and for $\eta=1$ 
(a soliton with $10^3$ atoms and width $2\xi=4\mu$m), it is clear that $L_{S}/L_{B}=2\xi \epsilon /B_{0}$.
This means that for sufficiently small values of the magnetic field gradient, e.g., for $\epsilon=5$ G/$\mu$m, 
the perturbation $R$ is of order $O(10^{-2})$. In such a case, we may employ the adiabatic perturbation theory for 
solitons \cite{kima}, to obtain the following equation of motion for the soliton center,
\begin{eqnarray} 
\frac{d^2x_{0}}{dt^2}=
-\frac{\partial V_{\rm eff}}{\partial x_{0}}, \,\,\,\,\,\, V_{\rm eff}(x_{0}) \equiv -\frac{1}{6} g^{2}(x_{0})
\label{eq_mot} 
\end{eqnarray} 

\begin{figure}[tbp]
\includegraphics[width=4.327cm]{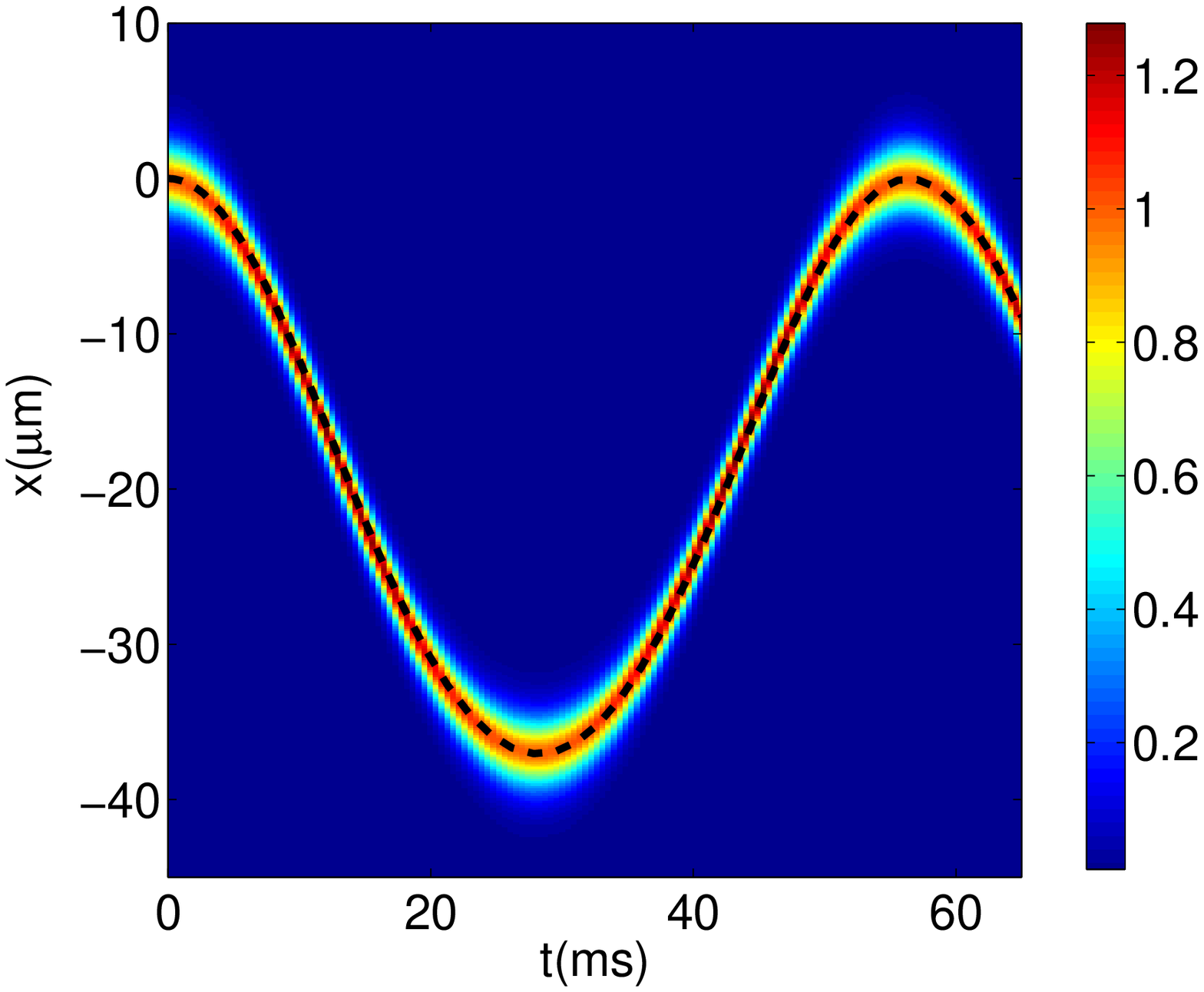}
\includegraphics[width=4.223cm]{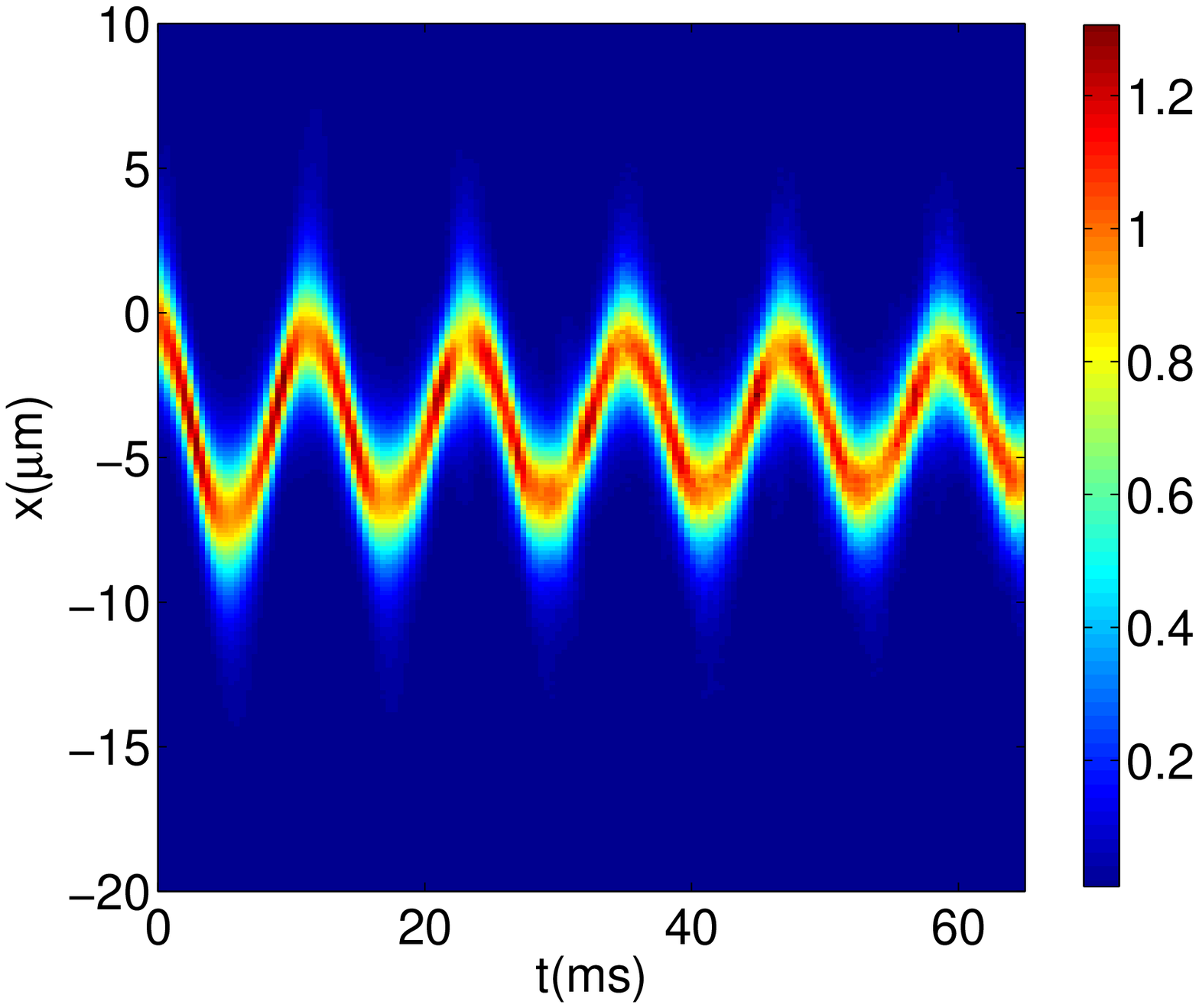}
\vskip -0.1in
\caption{(Color online) Spatio-temporal contour plot of the density of a bright matter wave soliton for $B_{0}=450 G$ 
and magnetic field gradients $\epsilon=5 G/\mu m$ (left panel) and $\epsilon=25 G/\mu m$ (right panel). 
The dashed line in the left panel corresponds to the analytical prediction according to Eq. (\ref{eq_mot}).}
\label{fig2}
\end{figure}

With $g(x)$ being proportional to $a(x)$, it is obvious (see Fig. \ref{fig1}) 
that 
the soliton ``feels'' a collision-induced effective confinement 
potential $V_{\rm eff}$,
{\it although there is no external trapping potential} [see Eq. (\ref{GP})].
It is then natural to expect that the collision-induced confinement
leads to oscillations of the soliton if it is displaced from the minimum of the confinement potential.
This has been verified by direct numerical integration of the GP Eq. (\ref{GP})
for $\epsilon=5$ G/$\mu$m and  initial condition $\psi={\rm sech}(x)$ 
(i.e., the soliton is initially at $x_{0}(0)=0$ where $B=B_{0}$). 
The result is shown in the left panel of Fig. 2, where 
the spatio-temporal contour plot of the soliton density is directly compared to the analytical 
prediction of Eq. (\ref{eq_mot}) (dashed line); 
the agreement between the two is excellent. 
We observe that the matter-wave is periodically compressed whenever it reaches the region of 
large scattering lengths ($x\approx -18\mu$m with $a \approx -0.23$nm), 
thus exhibiting a pronounced breathing behavior. 
This results in a spontaneous and 
robust {\it breathing behavior} of the matter-wave
soliton in the collisionally inhomogeneous environment.

For significantly larger field gradients, e.g., $\epsilon=25$ G/$\mu$m, 
the soliton width becomes comparable to the magnetic length scale $L_{B}$.
This means that the perturbation $R$ in Eq. (\ref{gpe1d_u}) is now of the order $L_{S}/L_{B}=O(10^{-1})$
and nonadiabatic effects are expected to be significant.  
The numerical integration of the GPE confirms this expectation
and reveals that although the soliton is still confined and performs corresponding oscillations, 
its evolution is nonadiabatic, i.e., emission of 
small amplitude wave radiation is observed. The corresponding results
are shown in the right panel of Fig. 2: Larger field gradients lead 
to oscillations with a higher frequency and smaller amplitude.

{\it Collision-induced transmission.} In our second setup we consider
the scattering of a bright matter-wave soliton (see Eq. (\ref{sol})
for $t=0$ and $\eta=1$) of an external potential barrier 
thereby comparing the results for homogeneous and 
inhomogeneous atomic interactions. An important quantity in this context is the transmission coefficient $T$. 

In order to compare the transmission in the above mentioned cases, 
the incoming and outgoing scattering environments should be identical. This means that the
scattering length should asymptotically, i.e. outside the range of the barrier, take on the
same values. We guarantee this by employing a localized inhomogeneous magnetic field of the form, 
\begin{eqnarray}
B(x)=\frac{1}{2}\left[\left(B_{1}+B_{2}\right)+ \left(B_{1}-B_{2}\right)\tanh(w(x-x_{\rm B})) \right],
\label{tanh} 
\end{eqnarray} 
where the parameters $w$ and $x_{\rm B}$ characterize the inverse width and location of the region of
inhomogeneity, respectively, while the field values $B_{1}$ and $B_{2}$ are chosen to
obtain equal scattering lengths $a(B_{1})=a(B_{2})$ sufficiently far from the barrier.
This value will also be used for the homogeneous case. In the case of $^{7}$Li  
it is straightforward to find the values $B_{1}$ and $B_{2}$, due to the convenient form of the function $a(B)$ 
in the considered range of field strengths (see also Fig.1). Here we use $B_{1}=450$ G and $B_{2}=265$ G 
for which $a=-0.182$ nm (other choices are, of course, equally possible and 
lead to similar results). We finally note that the above mentioned field configuration
can be realized by a multi-wire setup or a current density flowing in a half plane
augmented by a homogeneous bias field.

Our potential barrier is assumed to be of the form
$V_{\rm b}(x)=V_{0}{\rm sech}^{2}\left(\alpha(x-x_{\rm B})\right)$, 
where $V_{0}$, $\alpha^{-1}$ and $x_{\rm B}$ are the barrier's amplitude, width and location, respectively 
(note that the inhomogeneity is centered 
at the same position where the barrier is located, 
i.e., at $x=x_{\rm B}$). In the following we assume $V_{0}=1$ and $\alpha^{-1}=1/2$, i.e., 
the width of the barrier is half the soliton width, so as to avoid the classical Ehrenfest regime. 

The setup is illustrated in Fig. 3: The initial ($t=0$) 
form of the soliton, as well as its transmitted and reflected parts (for the inhomogeneous case 
at $t=6$ms) are depicted and labeled respectively. 
At the location of the barrier (shaded region), and for the collisionally inhomogeneous case,
there exists a local spatial change of the scattering length, which is obtained by inserting
the magnetic field in Eq. (\ref{tanh}) into the function $a(B)$ (utilizing the fifth-order polynomial
fit used in the previous section). The spatial dependence of the scattering length is also shown in Fig. 3; 
as discussed above, the scattering length takes on equal
values far from the barrier for both the homogeneous and inhomogeneous case.

\begin{figure}[tbp]
\includegraphics[width=6cm]{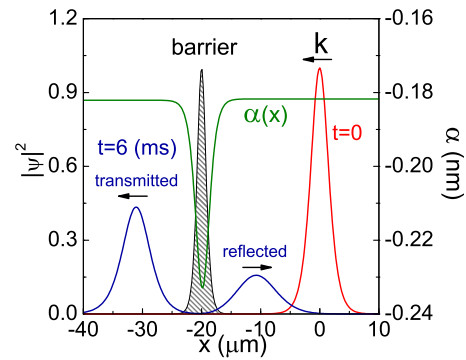}
\caption{
(Color online) 
The scattering of a soliton  initially placed at $x=0$ off a barrier 
(shaded area) located at 
$x_{\rm B}=-20\mu$m is shown; the transmitted and reflected parts of the 
soliton (accordingly labeled) are shown for $t=6$ms
for the inhomogeneous case. The spatial dependence of the scattering length is also shown. 
Note that for the homogeneous case $a=-0.182$nm.}
\label{tunnel2}
\end{figure}

\begin{figure}[tbp]
\includegraphics[width=4.275cm,height=4.1cm]{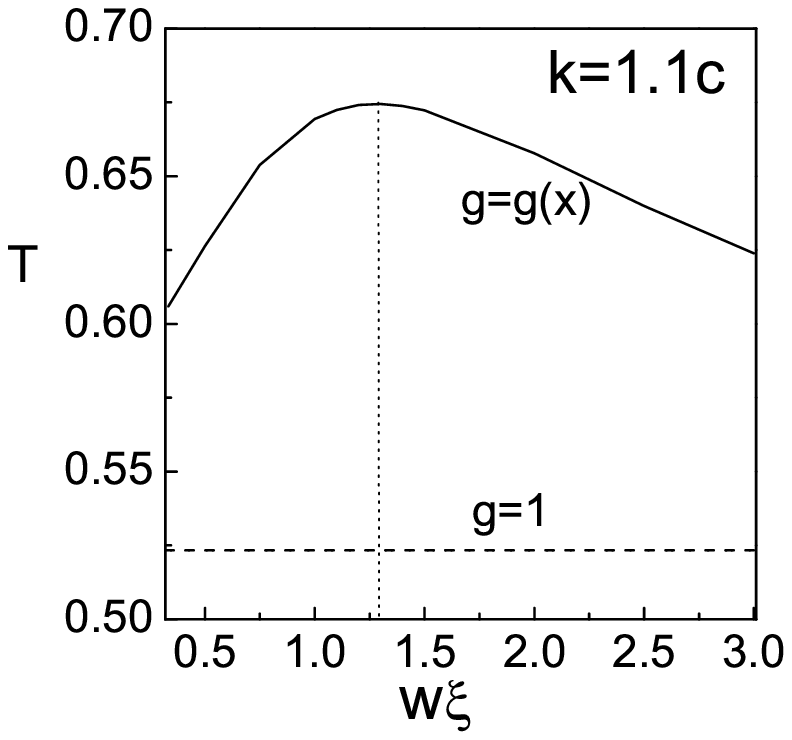}
\includegraphics[width=4.275cm,height=4cm]{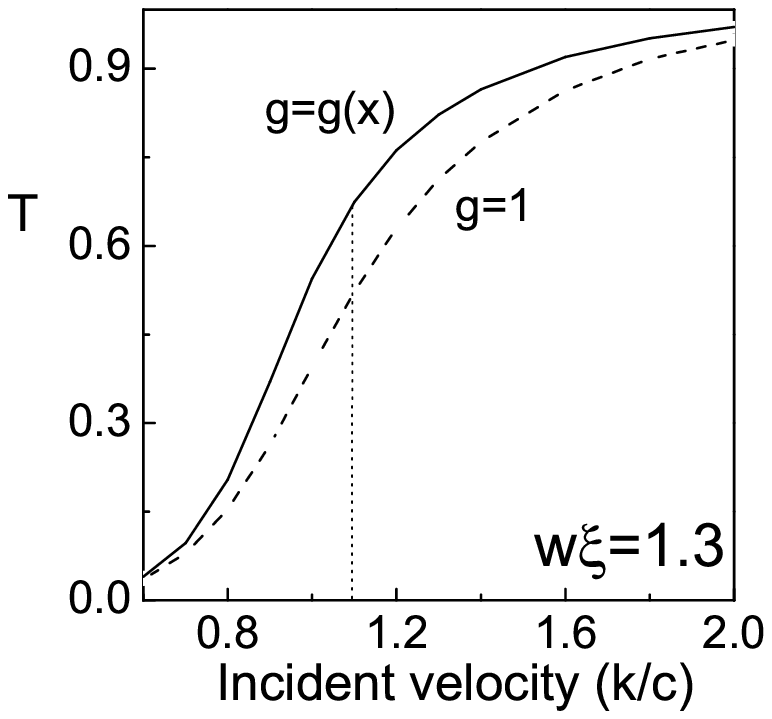}
\caption{
The transmission $T$ as a function of the inverse width $w$ of the inhomogeneity, in units of $\xi^{-1}$ (left panel), 
or the soliton's incident velocity $k$, in units of $c$ (right panel). Dashed and solid lines correspond to the 
collisionally homogeneous and inhomogeneous cases, respectively.
}
\label{tunnel}
\end{figure}

We have numerically integrated the GPE incorporating the potential barrier
term   
$V_{\rm b}(x) \psi(x,t)$ to determine the transmission coefficient $T$. 
The results presenting
 $T$ as a function of the width of the inhomogeneity (left panel) 
and the soliton's incident velocity (right panel) are shown in Fig. 
\ref{tunnel}.
Generally, for a fixed soliton velocity, or width of the inhomogeneity, the 
transmission $T$ in the inhomogeneous case is always larger than the one in
the homogeneous case. Particularly, for $k=1.1c$ and $w\xi=1.3$, the relative
difference of the transmission $T$ for the two cases becomes maximal, being 
$\approx 15\%$. 
This result clearly demonstrates that the 
transmission of a matter-wave soliton 
may be enhanced in the presence of a 
spatially dependent collisional interaction.

{\it Conclusions.} We have explored the dynamics of bright matter-wave solitons 
subject to a spatially varying nonlinearity, which can be realized by means of an external inhomogeneous
magnetic field on top of a bias field in order to be close to a region of strong changes of the scattering
length, such as a Feshbach resonance. It was demonstrated that a confinement (trapping) of the matter-wave
can be achieved solely on basis of the spatial change of the collisional interaction, 
i.e., without the presence of an external trapping potential thereby creating a 
breathing matter-wave state in the collisionally inhomogeneous environment.
In an adiabatic regime, such a state could be well described within the 
realm of soliton perturbation theory, while for abrupt variations
of the scattering length, radiative emissions are non-trivial resulting
in energy losses and hence shorter period oscillations.
Using a localized spatial variation of the scattering
length, we have shown that the transmission of matter-wave solitons 
through a barrier can be enhanced
by suitably manipulating the collisional properties of the condensate in
the vicinity of the potential barrier.
Collisionally inhomogeneous environments therefore 
hold considerable promise in the effort to
control and manipulate matter-waves in experimental
applications. Interesting future directions may involve combining
this type of effective potential with optical potentials in order
to examine transmission properties and symmetry breaking or
nonlinear trapping phenomena similarly to the recent works of
\cite{prlpla}. An interesting variation on these themes that
the present setting offers is, among others, that the effective
potential is asymmetric and can hence lead to a modified dynamical
picture \cite{kevkap} in comparison to the symmetric potential
case \cite{prlpla}. Such studies are currently in progress and
will be reported in future publications.

{\bf Acknowledgements.} This work was supported by 
the ``A.S. Onasis'' Public Benefit Foundation (GT),
the Special Research Account of Athens University (GT, DJF), 
as well as NSF-DMS-0204585, NSF-CAREER, NSF-DMS-0505063 
and the Eppley Foundation for Research (PGK).


\end{document}